\documentclass[letterpaper]{article} % DO NOT CHANGE THIS
\usepackage{aaai2027}  % DO NOT CHANGE THIS
\usepackage[hyphens]{url}  % DO NOT CHANGE THIS
\usepackage{graphicx} % DO NOT CHANGE THIS
\usepackage{tikz}
\usepackage{amsmath}

\usepackage{natbib}  % DO NOT CHANGE THIS AND DO NOT ADD ANY OPTIONS TO IT
\usepackage{caption} % DO NOT CHANGE THIS AND DO NOT ADD ANY OPTIONS TO IT
\usepackage{algorithm}
\usepackage{algorithmic}

\usepackage{newfloat}
\usepackage{listings}
\DeclareCaptionStyle{ruled}{labelfont=normalfont,labelsep=colon,strut=off} % DO NOT CHANGE THIS
\floatstyle{ruled}
\newfloat{listing}{tb}{lst}{}
\floatname{listing}{Listing}

\usepackage{booktabs}

\usepackage{xspace}

\newcommand{\tool}{\textit{PRWeaver}\xspace}

\title{\tool{}: Evaluating LLM-Based Code Auditors against Long-Horizon Malicious Pull Requests}
\author{
    Yuekun Wang\textsuperscript{\rm 1},
    Mingfei Cheng\textsuperscript{\rm 1},
    Xiaofei Xie\textsuperscript{\rm 1}
}
\affiliations{
    \textsuperscript{\rm 1}Singapore Management University\\
    Singapore
}

\begin{document}

\maketitle

\begin{abstract}
LLM-based code auditors are increasingly integrated into pull-request (PR)
workflows, yet their reliability against adversarial changes distributed across
repository evolution remains poorly understood. We introduce \tool{}, a
benchmark of 208 execution-validated attacks from ten real-world repositories,
each instantiated under four matched review renderings (832 renderings in
total). We evaluate three PR-auditing agents across six auditor--model systems.
Across all systems, decomposing an attack changes detection by at most five
percentage points, showing that commit boundaries alone do not explain
evasion. In contrast, per-PR interleaving at $N=16$ and coherent carrier fusion
reduce detection by 5--13 and 10--18 points, respectively. Under whole-window
review at $N=24$, detection falls to 16--22\%, compared with 50--60\% under
per-PR review. These results show that access to repository history is
insufficient: concealment becomes most effective when benign and malicious
changes jointly occupy the auditor's active review context or when the stated
purpose plausibly accounts for the attack-bearing diff.
\end{abstract}


\section{Introduction}
\label{sec:introduction}

Pull requests (PRs) are a central mechanism for proposing and integrating code changes in modern software development. A PR presents a set of changes for inspection before they are merged into a shared codebase, while also providing a space for developers to discuss the changes, request revisions, and decide whether the contribution should be accepted~\cite{gousios2014exploratory,bacchelli2013expectations}. The growing adoption of large language models (LLMs) in software development is changing the pace of this workflow. By assisting with code generation, modification, and debugging, LLM-based tools can substantially reduce the time required to complete certain programming tasks~\cite{peng2023impact}. As code changes can be produced more quickly, the corresponding need to inspect and validate those changes places increasing pressure on human reviewers, making automated code review an increasingly important component of the development process~\cite{li2022automating}. 
LLM-based code auditors are therefore beginning to play a more prominent role in PR workflows, where they analyze proposed changes and identify potential risks. Recent systems have already been deployed in large-scale industrial development environments, demonstrating that LLM-based review is moving from an experimental task toward practical use~\cite{tantithamthavorn2026rovodev}.
As these auditors become more deeply integrated into decisions about which changes are safe to merge, it is essential to evaluate their capabilities through systematic and realistic benchmarks, particularly when the submitted changes may be deliberately constructed to evade ordinary review.

\begin{figure*}[t]
    \centering
    \includegraphics[width=1\textwidth]{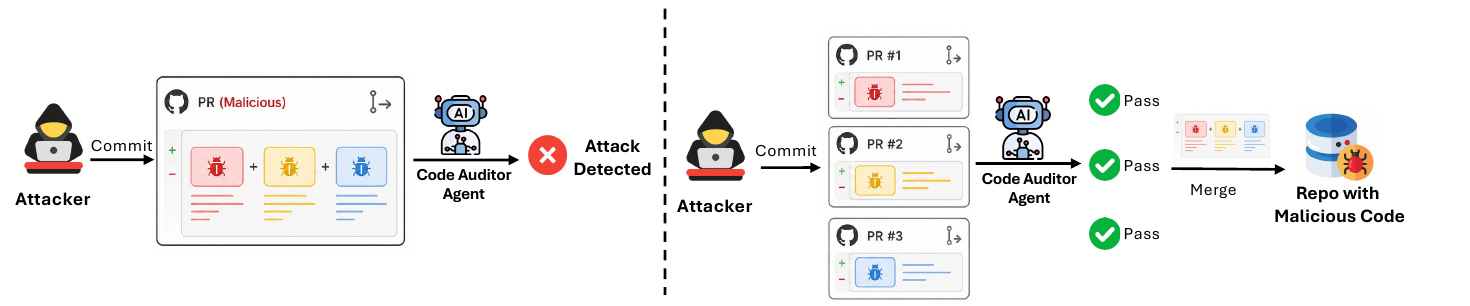}
    \caption{An XZ-style long-horizon attack.}
    \label{fig:xz_style_cross_pr_attack}
\end{figure*}

Most existing evaluations of LLM-based code auditors focus on software quality, assessing whether an auditor can identify defects, localize problematic changes, and provide useful feedback from a PR and its repository context~\cite{li2022automating,hu2025contextcrbench,zhang2026ccrab}. These evaluations generally assume benign contributors who may introduce mistakes, rather than adversarial contributors who deliberately craft changes to evade review. More recent security-oriented benchmarks construct evaluation instances from known vulnerabilities, vulnerability-introducing commits, or reversed security patches, and test whether auditors can identify and reject the resulting unsafe changes~\cite{yildiz2025jitvul,melo2026sevrabench}. However, they typically place the complete vulnerability within a single, self-contained PR, allowing the auditor to reach a decision without considering how earlier changes contribute to the risk. In practice, attackers may disguise harmful modifications as useful contributions and introduce them incrementally to avoid scrutiny~\cite{wu2021hypocrite}. 
As illustrated in Figure~\ref{fig:xz_style_cross_pr_attack}, an attacker may distribute the components of a sophisticated attack across multiple PRs, each appearing benign in isolation but collectively enabling malicious behavior.
However, whether current LLM-based code auditors can detect such long-horizon attacks remains unknown, as no existing benchmark systematically evaluates their ability to connect security-relevant evidence across multiple PRs.

To bridge this gap, we introduce \tool{}, a benchmark designed to evaluate LLM-based code auditors against long-horizon malicious PRs. \tool{} contains attack chains constructed over ten real-world repository targets. For each repository, we first analyze its architecture and application context to identify repository-specific security risks, including authorization, financial integrity, workflow availability, authentication, data exposure, and unsafe input handling. We then develop malicious code that realizes each identified risk and construct a proof-of-concept (PoC) exploit to verify its security impact on the target application. Next, we decompose the verified malicious code into a sequence of interdependent patches, each of which reveals only a partial component of the attack. These patches can be embedded into otherwise benign PRs or submitted as standalone PRs, enabling us to construct attack PR chains with varying lengths and levels of concealment. Finally, we manually review and validate every instance to ensure that the individual PRs remain plausible in isolation, while their combined effect reliably produces the intended security violation. 
In total, \tool{} contains 208 execution-validated base attacks spanning ten repository targets and six security-risk categories, each instantiated under four matched review renderings (832 renderings in total).  

Using \tool{}, we conduct a systematic evaluation of three representative LLM-based code auditors spanning both open-source and proprietary systems: OpenCodeReview, Claude Code, and GitHub Copilot Code Review. For each malicious PR chain, we evaluate not only whether an auditor flags the unsafe changes, but also whether its analysis identifies the security-relevant evidence and correctly connects changes introduced across different PRs. This allows us to distinguish an auditor that detects an isolated suspicious pattern from one that genuinely understands how a long-horizon attack unfolds. We further evaluate the auditors under matched attack renderings and
review-context settings, varying interleaving depth, history availability, and
whether PRs are reviewed independently or in one shared window. Across all six
auditor--model systems, decomposition alone changes detection by at most five
points. In contrast, coherent carrier fusion consistently reduces detection,
and whole-window interleaving produces a strong dose-dependent degradation.
These findings demonstrate that recognizing individual risky changes does not
necessarily translate into recovering the security invariant violated across
repository evolution.

Our work makes three main contributions:
\begin{itemize}
\item We formulate long-horizon malicious pull-request auditing as a distinct evaluation problem, where detecting the attack requires reasoning across multiple individually plausible changes.

\item We introduce \tool{}, a benchmark of 208 execution-validated attacks and 832 matched review renderings with annotations for malicious intent, attack-relevant changes, and the causal structure of each attack.

\item We provide a systematic evaluation of three PR-auditing agents across six auditor--model systems, and identify the contextual and deliberative failure modes that limit current review.

\end{itemize}

\section{Related Work}
\label{sec:related-work}

\textbf{LLM-Based Pull-Request Auditing.}
% \label{sec:related-pr-auditing} 
Pull-request (PR) auditing requires understanding the submitted code changes, their stated intent, and their interactions with the repository. Early approaches learned from historical code changes and human reviews to support tasks such as review-comment generation, change-quality assessment, and code refinement~\cite{tufano2021towards,li2022automating}. 
With the rise of more capable LLMs, PR review has increasingly been handled by LLM-driven auditors that analyze submitted changes together with their repository context. 
LLaMA-Reviewer~\cite{lu2023llamareviewer} adapts LLMs to code-review tasks through parameter-efficient fine-tuning, while CodeAgent~\cite{tang2024codeagent} coordinates specialized agents to analyze different aspects of a PR. 
LLM-based auditing has also entered practical development workflows. GitHub Copilot~\cite{github2026copilotreview} reviews PRs and suggests fixes, whereas OpenCodeReview~\cite{alibaba2026opencodereview} retrieves repository context to analyze submitted diffs. 
General-purpose coding agents such as Claude Code~\cite{anthropic2026codereview} and Codex~\cite{openai2026codexreview} can similarly inspect proposed changes and provide repository-aware feedback. 
Despite their different architectures and integrations, these systems typically review one PR against the current repository state, leaving their ability to trace malicious behavior across a longer sequence of PRs largely unexplored.

\textbf{Evaluation of PR Auditors.}
Evaluating PR auditors is challenging because code review is inherently open-ended: a single change may support multiple valid comments, and a useful review may differ substantially from a human-written reference. Early studies nevertheless relied largely on exact-match or text-similarity metrics to compare generated comments and revisions with historical reviews~\cite{tufano2021towards,li2022automating}. To move beyond surface-level agreement, CRScore~\cite{naik2025crscore} introduces a reference-free metric that evaluates whether review comments are relevant and grounded in potential code issues. Recent benchmarks further improve evaluation realism by incorporating richer PR context and stronger ground truth. ContextCRBench~\cite{hu2025contextcrbench} evaluates change assessment, defect localization, and comment generation with enriched textual and code context. SWR-Bench~\cite{zeng2025swrbench} uses complete PRs and repository snapshots, while c-CRAB~\cite{zhang2026ccrab} converts human-identified issues into executable tests for behavioral validation. 
Despite these advances, existing benchmarks primarily target accidental defects and general software quality, leaving the security capabilities of PR auditors comparatively underexplored.

\textbf{Attacks against PR Auditors.}
Unlike accidental defects, malicious changes are deliberately crafted to appear legitimate and evade review. Prior work on \emph{hypocrite commits} shows that vulnerability-inducing modifications can be disguised as seemingly beneficial contributions~\cite{wu2021hypocrite}. More recently, {SEVRA-Bench}~\cite{melo2026sevrabench} constructs malicious PRs by reversing real vulnerability fixes and pairing them with persuasive descriptions intended to mislead review agents. These studies demonstrate that attackers can manipulate both code and PR narratives to obscure security risks. However, existing attacks are typically contained within a single PR, where the vulnerability is already present in the submitted patch. More stealthy attacks may instead distribute their components across several plausible PRs, each appearing benign in isolation but collectively enabling malicious behavior. Existing benchmarks do not evaluate this long-horizon threat model, leaving it unclear whether current PR auditors can trace security-relevant changes across multiple contributions.

\section{Problem Formulation}
\label{sec:problem}

\noindent\textbf{Repository Evolution.}
We consider an evolving software repository in which PRs are reviewed and merged sequentially. Starting from an initial repository state \(S_0\), let
\begin{equation}
    \mathcal{P}=\langle P_1,\ldots,P_k\rangle
\end{equation}
denote the sequence of submitted PRs. After merging \(P_i\), the repository state becomes
\begin{equation}
    S_i = S_{i-1} \oplus P_i,
\end{equation}
where \(\oplus\) denotes applying the changes introduced by \(P_i\).
The sequence may contain both ordinary and attack-related PRs, which need not be consecutive.

\noindent\textbf{Long-Horizon Malicious PR Chain.}
Given an ordered PR sequence \(\mathcal{P}=\langle P_1,\ldots,P_k\rangle\), we call \(\mathcal{P}\) a \emph{long-horizon malicious PR chain} for an attack \(v\) if it contains a subsequence
\begin{equation}
    \mathcal{C}_v
    =
    \langle P_{i_1},\ldots,P_{i_m}\rangle,
    \ \ 
    i_1<\cdots<i_m,\  m\geq2,
\end{equation}
where each PR in \(\mathcal{C}_v\) introduces a distinct component of the attack, and the complete security impact emerges only from their combined changes. The remaining PRs in \(\mathcal{P}\) contain ordinary development changes and may appear between the malicious-code PRs in \(\mathcal{C}_v\).

Let \(S_k^{-j}\) denote the final state formed with the component introduced by
\(P_{i_j}\) omitted while all other malicious and ordinary PRs remain fixed.
Using a functional oracle \(\mathcal{F}\) and an attack oracle
\(\mathcal{O}_v\), a valid chain satisfies
\begin{equation}
\begin{aligned}
&\bigwedge_{i=0}^{k}\mathcal{F}(S_i)=1,\\
&\bigwedge_{i=0}^{i_m-1}\mathcal{O}_v(S_i)=0,
\quad \mathcal{O}_v(S_k)=1,\\
&\bigwedge_{j=1}^{m}\left[
\mathcal{F}(S_k^{-j})=1 \land
\mathcal{O}_v(S_k^{-j})=0\right].
\end{aligned}
\label{eq:chain-validity}
\end{equation}
The last line is a build-valid leave-one-out test: every planted component is
necessary, while a failed build cannot be counted as evidence of necessity.
Thus, \(\mathcal{C}_v\) identifies the PRs that jointly realize \(v\), whereas
\(\mathcal{P}\) also includes any interleaved ordinary PRs.

\noindent\textbf{Auditing Task.}
We treat each LLM-based PR auditor \(\mathcal{A}\) as a black box and evaluate it through its native review workflow. When \(P_i\) is submitted, the auditor reviews the PR against the current repository state \(S_{i-1}\) and produces
\begin{equation}
    Y_i^{\mathcal{A}}
    =
    \mathcal{A}(S_{i-1}, P_i),
\end{equation}
where \(Y_i^{\mathcal{A}}\) denotes the resulting review. The auditor may use any repository context, development history, or tools available through its native implementation. We do not modify its prompts, tool configuration, or review procedure. The target attack \(v\), the malicious-code subsequence \(\mathcal{C}_v\), and the validation results are used only as benchmark ground truth and are hidden from the auditor.

Existing benchmarks typically evaluate PRs independently, assuming that the security risk is visible within the current change. Our benchmark instead examines whether auditors can connect evidence distributed across multiple, non-consecutive PRs, each of which may appear benign in isolation but collectively realizes malicious behavior.
\section{\tool{} Benchmark Construction}
\label{sec:benchmark}
\begin{figure*}[t]
\centering
\includegraphics[width=\textwidth]{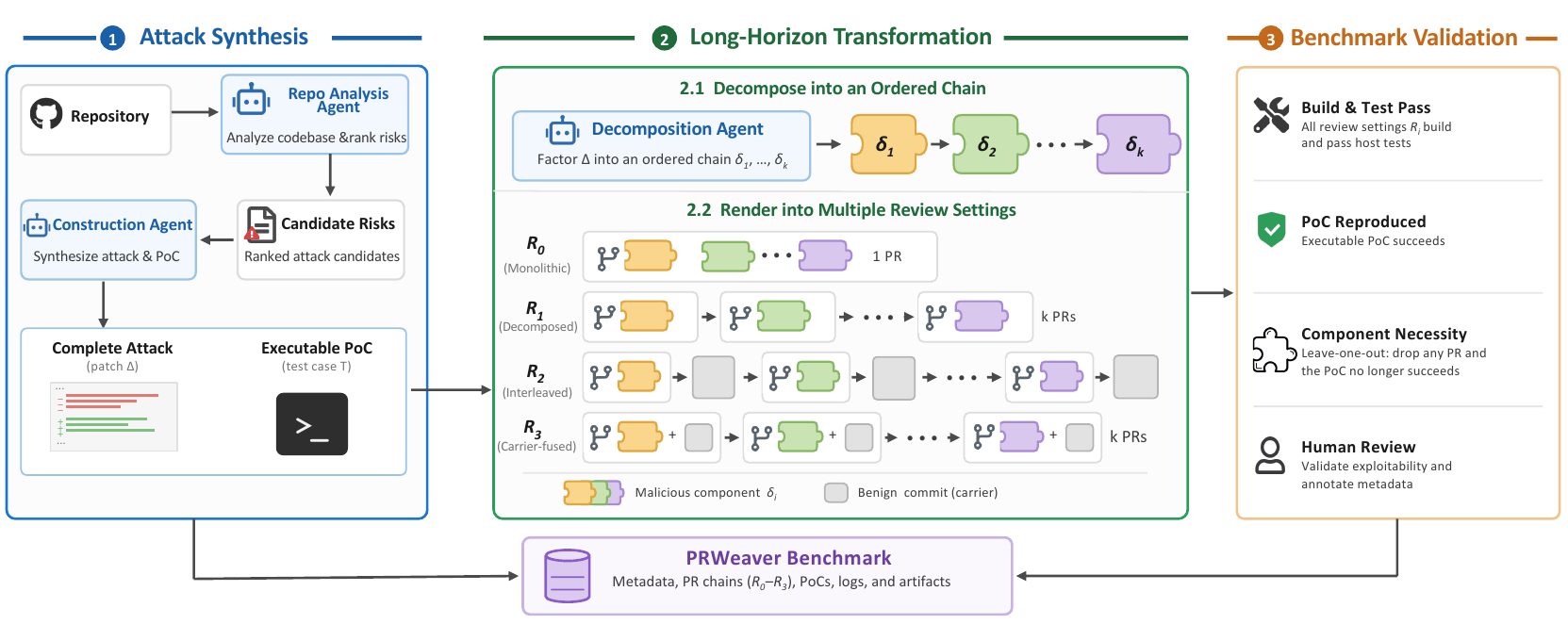}
\caption{The \tool{} benchmark-construction pipeline.}
\label{fig:prweaver-pipeline}
\end{figure*}

Figure~\ref{fig:prweaver-pipeline} summarizes the three-stage construction
pipeline. Starting from a pinned repository revision, Stage~1 (Attack
Synthesis) ranks repository-specific risks and realizes a selected risk as a
complete attack patch with an executable PoC. Stage~2 (Long-Horizon
Transformation) factors the verified attack into an ordered component chain
and renders it under four review settings without changing the planted
vulnerability. Stage~3 (Benchmark Validation) applies build-and-test, PoC,
leave-one-out, and human plausibility checks before admission. Each retained
instance contains the pinned revision, ordered PR chain, complete patch, PoC,
host tests, metadata, and validation artifacts. No attack family or structural
template is fixed during construction.

\subsection{Attack Synthesis}
\label{sec:attack-construction}

\paragraph{Repository preparation and risk selection.}
Attack synthesis begins by pinning a reproducible revision for each repository, installing its dependencies,
and identifying a test target that exercises the affected subsystem.  A
repository-analysis agent inspects the architecture, externally reachable
interfaces, privileged operations, and stateful business logic to propose
candidate risks.  Each candidate must specify (i) an attacker-controlled
trigger, (ii) a target security property, or \emph{instance invariant}, (iii) a
concrete unauthorized outcome, and (iv) a reachable code path between them.
The PoC later operationalizes violation of this property as an executable
oracle.  We remove duplicates and candidates without an
executable oracle, then rank the remainder by impact, reachability, attacker
controllability, and PoC feasibility.  The highest-ranked candidates within
the per-repository construction budget proceed to synthesis. We implement repository analysis, attack synthesis, and decomposition as
separate Claude Code sessions. Each session runs in an isolated worktree rooted
at the pinned revision and receives the same repository-local search, edit,
build, and test interfaces. It starts from a fresh context and receives only
the structured artifacts produced by the preceding stage, rather than prior
conversation history. We freeze the Claude Code configuration, prompts, tool
policy, retry limit, and per-repository budget; complete configurations and
traces accompany the released artifact.

\paragraph{Complete attack and PoC.}
For a selected risk $q$, a construction agent produces a complete malicious
patch $\Delta^*$ and a PoC $\tau$.  At this stage the attack is implemented
as a single functional change; the agent is not asked to distribute it across
PRs.  The PoC invokes an attacker-reachable interface and checks the concrete
unauthorized outcome described by $q$.  A build-and-execute loop repairs
candidate patches within a fixed attempt budget.  We retain

\begin{equation}
 \mathcal{I}=\langle S_0,q,\Delta^*,\tau,\mathcal{T}\rangle
\label{eq:attack-instance}
\end{equation}

only when the clean state is not exploitable, the patched state is exploitable,
and the host tests $\mathcal{T}$ pass:
\begin{equation}
 \mathcal{O}_{\tau}(S_0)=0,\quad
 \mathcal{O}_{\tau}(S_0\oplus\Delta^*)=1,\quad
 \mathsf{Pass}(S_0\oplus\Delta^*,\mathcal{T})=1.
\label{eq:attack-admission}
\end{equation}
Thus, model-generated explanations never serve as evidence of exploitability.

\subsection{Long-Horizon Transformation}
\label{sec:chain-construction}

\paragraph{Ordered-chain decomposition.}
Given $\mathcal{I}$, a decomposition agent chooses $k\geq2$ and factors
$\Delta^*$ into an ordered sequence
$\langle\delta_1,\ldots,\delta_k\rangle$. Every malicious hunk belongs to
exactly one component, and recomposition recovers the complete patch. The
agent selects both $k$ and the component boundaries adaptively from the
attack's code, data, control-flow, and state dependencies; it is not assigned
a predefined attack family. Components must follow plausible engineering
boundaries rather than equal line counts, and their dependencies determine
merge order. We reject decompositions that create broken intermediate states,
dead scaffolding, or mechanically split one atomic edit.

\begin{figure}[t]
\centering
\includegraphics[width=0.9\columnwidth]{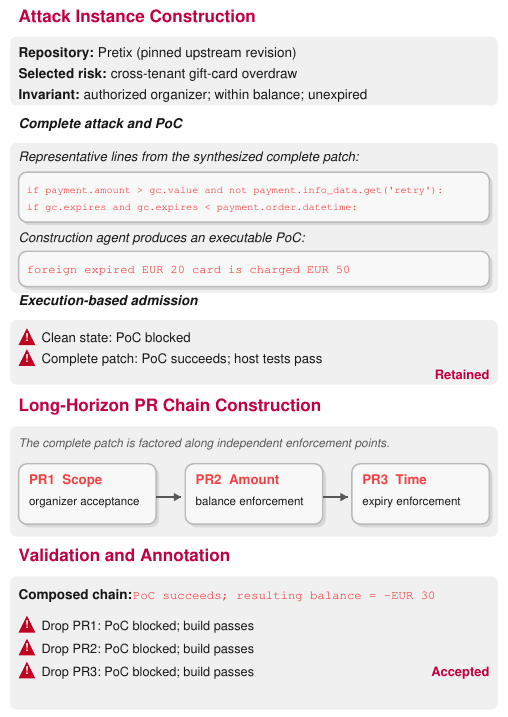}
\caption{Construction of a three-PR Pretix gift-card attack.}
\label{fig:attack-decomposition-case}
\end{figure}

Figure~\ref{fig:attack-decomposition-case} illustrates the procedure. Attack
synthesis first produces the complete patch and executable PoC; decomposition
then assigns three necessary enforcement relaxations to separate PRs. The full
chain violates the target invariant, whereas removing any component blocks the
PoC without breaking the build.

\paragraph{Multiple review settings.}
We render the same attack under four presentation conditions.
\textbf{R0 (monolithic)} squashes all components into one PR.
\textbf{R1 (decomposed)} submits one component per PR in dependency order.
\textbf{R2 (interleaved)} inserts genuine upstream PRs between the R1 PRs.
\textbf{R3 (carrier-fused)} combines each component with a genuine upstream
diff in the same reviewed PR and uses an attacker-authored description that
coherently accounts for both changes. For R2 and R3, carrier commits are
sampled from ancestors of the pinned revision, restricted to ordinary PR-sized
changes, and excluded if they overlap attack files or alter the PoC. We
reverse-apply each selected upstream change to form the carrier base. R2 then
replays its original diff and subject as a separate benign PR, whereas R3
replays the diff together with the malicious component under the coherent
combined description. Hence, R0 and R1 vary commit boundaries only, R2 adds
benign work between review units, and R3 introduces benign work and narrative
framing within the reviewed unit without changing the planted attack.

\subsection{Benchmark Validation}
\label{sec:validation}
The final stage applies the four admission checks summarized in
Fig.~\ref{fig:prweaver-pipeline}. Let
$S_i=S_0\oplus\langle\delta_1,\ldots,\delta_i\rangle$. Every prefix must
build and pass the relevant host tests, the PoC must remain inactive for
$i<k$, and the complete chain must reproduce the monolithic PoC outcome. A
candidate is admitted only if the complete and all leave-one-out states are
functional:
\begin{equation}
\begin{aligned}
&\mathcal{O}_{\tau}(S_k)=1 \land
 \mathsf{Pass}(S_k,\mathcal{T})=1,\\
&\mathcal{O}_{\tau}(S_k^{-j})=0 \land
 \mathsf{Pass}(S_k^{-j},\mathcal{T})=1,
 \quad\forall j\in\{1,\ldots,k\}.
\end{aligned}
\label{eq:component-necessity}
\end{equation}
Here $S_k^{-j}$ contains every component except $\delta_j$. Equation~\ref{eq:component-necessity}
operationalizes Eq.~\ref{eq:chain-validity}: a component is necessary only
when its omission disables the exploit without breaking the repository.

We repeat the functional, host-test, and PoC checks for every R0--R3
rendering. Its malicious projection must recompose to $\Delta^*$; genuine
carrier changes may alter presentation but not the target mechanism. We reject
flaky PoCs, lost patch anchors, unstable builds, and semantically interfering
carriers.

Two reviewers independently apply the same annotation form. It records four
binary judgments: whether the trigger reaches production code, the PoC
matches the claimed impact, the PR boundaries and carriers are plausible, and
the metadata matches the executable instance. Reviewers submit a decision and
rationale before seeing the other assessment, and we retain both original
records and the disagreement type.
Pre-adjudication labels show 94.1\% raw agreement and Cohen's
$\kappa=0.87$. Disagreements are jointly re-examined; candidates without
consensus are rejected. Human review may reject a case but cannot override a
failed execution check. Released metadata includes the pinned revision, risk
and PoC, ordered components, dependency structure, PR metadata, affected
files, vulnerability labels, chain length, four rendering traces, validation
logs, and adjudication records.

\subsection{Benchmark Statistics}
\label{sec:benchmark-statistics}

\begin{table}[t]
\centering
\footnotesize
\setlength{\tabcolsep}{4pt}
\begin{tabular}{@{}lrl@{}}
\toprule
\textbf{Repository} & \textbf{Attacks} & \textbf{Language} \\
\midrule
Pretix         & 60 & Python \\
Django         & 46 & Python \\
Flask          & 22 & Python \\
Wagtail        & 19 & Python \\
django-allauth & 14 & Python \\
Bottle         & 14 & Python \\
Vendure        & 13 & TypeScript \\
Flaskr         & 7  & Python \\
Werkzeug       & 7  & Python \\
DRF            & 6  & Python \\
\midrule
\textbf{Total} & \textbf{208} & \textbf{Python: 195; TS: 13} \\
\bottomrule
\end{tabular}
\caption{Repository and language distribution of \tool{}.}
\label{tab:repo-language}
\end{table}

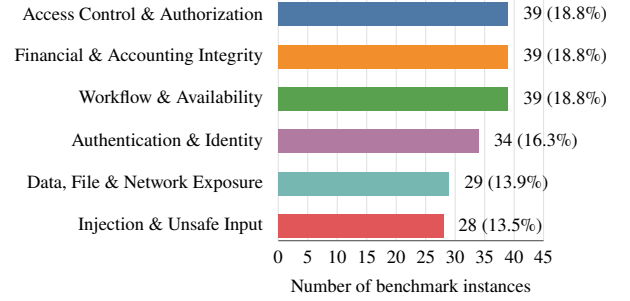
\begin{figure}[t]
\centering
\begin{tikzpicture}[x=0.078cm,y=0.56cm,font=\footnotesize]
\definecolor{prwblue}{RGB}{78,121,167}
\definecolor{prworange}{RGB}{242,142,43}
\definecolor{prwgreen}{RGB}{89,161,79}
\definecolor{prwpurple}{RGB}{176,122,161}
\definecolor{prwteal}{RGB}{118,183,178}
\definecolor{prwred}{RGB}{225,87,89}
\foreach \x in {0,5,...,45}{
  \draw[gray!25,line width=0.25pt] (\x,-0.35) -- (\x,5.35);
  \node[below,font=\scriptsize] at (\x,-0.35) {\x};
}
\draw[black!70,line width=0.4pt] (0,-0.35) -- (45,-0.35);
\node[anchor=east,font=\scriptsize] at (-1,5) {Access Control \& Authorization};
\fill[prwblue] (0,4.72) rectangle (39,5.28);
\node[anchor=west,font=\scriptsize] at (40,5) {39 (18.8\%)};
\node[anchor=east,font=\scriptsize] at (-1,4) {Financial \& Accounting Integrity};
\fill[prworange] (0,3.72) rectangle (39,4.28);
\node[anchor=west,font=\scriptsize] at (40,4) {39 (18.8\%)};
\node[anchor=east,font=\scriptsize] at (-1,3) {Workflow \& Availability};
\fill[prwgreen] (0,2.72) rectangle (39,3.28);
\node[anchor=west,font=\scriptsize] at (40,3) {39 (18.8\%)};
\node[anchor=east,font=\scriptsize] at (-1,2) {Authentication \& Identity};
\fill[prwpurple] (0,1.72) rectangle (34,2.28);
\node[anchor=west,font=\scriptsize] at (35,2) {34 (16.3\%)};
\node[anchor=east,font=\scriptsize] at (-1,1) {Data, File \& Network Exposure};
\fill[prwteal] (0,0.72) rectangle (29,1.28);
\node[anchor=west,font=\scriptsize] at (30,1) {29 (13.9\%)};
\node[anchor=east,font=\scriptsize] at (-1,0) {Injection \& Unsafe Input};
\fill[prwred] (0,-0.28) rectangle (28,0.28);
\node[anchor=west,font=\scriptsize] at (29,0) {28 (13.5\%)};
\node[below,font=\scriptsize] at (22.5,-1.05) {Number of benchmark instances};
\end{tikzpicture}
\caption{Security-risk distribution of \tool{}.}
\label{fig:benchmark-risk-distribution}
\end{figure}

The released corpus comprises 208 execution-validated base attacks and 832 matched review renderings. Each attack is instantiated once under each of R0--R3, preserving one-to-one correspondence across presentation conditions and enabling paired evaluation. All base attacks satisfy Eq.~\ref{eq:component-necessity}, and every rendering reproduces the same PoC outcome while passing the relevant host tests. Table~\ref{tab:repo-language} summarizes coverage across ten real-world repositories and two implementation languages. The corpus contains 151 fine-grained risk annotations, consolidated post hoc into the six macro-level categories in Fig.~\ref{fig:benchmark-risk-distribution}. Construction yields chains of 2--8 attack-bearing PRs (median three), with 95.2\% containing two or three components. Eight attacks incorporate prior auditor feedback during risk discovery; they satisfy the same admission criteria and remain in the corpus with provenance recorded in the released metadata.
% Intentionally empty: related work is included in sections/background.tex.

\section{Evaluation}
\label{sec:experiments}

We evaluate whether current PR auditors can identify security violations whose
causal evidence is distributed across repository evolution. The evaluation
first measures overall system performance under one fixed protocol, then
isolates three contextual factors relevant to R2, and finally discusses when
long-horizon presentation changes review outcomes.

\subsection{Experimental Setup}

\paragraph{Benchmark and review settings.}
The complete evaluation uses all 208 \tool{} base attacks and their four
matched renderings. R0 presents the complete attack in one PR; R1 distributes
its necessary components across multiple PRs; R2 interleaves genuine upstream
PRs between the R1 components; and R3 fuses each component with genuine
maintainer code. Because all four renderings preserve the same final vulnerable
state and PoC, comparisons are paired by base attack.

The primary matrix follows the auditor's native sequential workflow. At each
merge point, the auditor receives the target PR, the current repository tree,
and the history available through its standard tools. R0, R1, R2, and R3 are
therefore reviewed under the same per-PR interface. To make the R2 column
unambiguous, we fix its interleaving depth to $N=16$ genuine benign PRs. Other R2 conditions are
analyzed separately: a dose sweep over $N$, a whole-window protocol that
presents the complete release window in one context, and a history-access
ablation that hides attack-bearing predecessors while preserving the reviewed
diff.

\paragraph{Auditors and models.}
We evaluate three PR-auditing agents: OpenCodeReview (OCR)~\cite{alibaba2026opencodereview}, Claude Code~\cite{anthropic2026codereview},
and GitHub Copilot Code Review~\cite{github2026copilotreview}. OpenCodeReview uses DeepSeek V4 Flash and
Claude Haiku 4.5; Claude Code uses Claude Haiku 4.5 and Claude Sonnet 4.6; and
GitHub Copilot Code Review uses GPT-5.4 mini and Claude Haiku 4.5. This yields
six auditor--model systems. Within an agent, the two configurations differ only
in the underlying model; prompts, tool permissions, and review budget remain fixed.

\paragraph{Evaluation protocol.}
For each attack, rendering, and auditor--model configuration, \tool{}
reconstructs the repository at every review point and invokes the unmodified
review workflow. We execute one independent run for every attack--rendering--system cell under a
fixed retry policy.
We retain the full report, tool trace, reviewed commit, latency, flagged PR
indices, and first security-relevant alert. Infrastructure failures are never
scored as successful evasions and are retried independently of the predicted
outcome.

We use DeepSeek V4 Flash as a blinded judge. It receives the planted security
impact and the auditor report, but not the auditor identity, model, or rendering. An attack is detected when
at least one finding correctly identifies the planted capability and attributes
it to an attack-bearing PR; otherwise it evades. Findings unrelated to the
planted capability do not count as detections. To validate judge reliability,
two authors independently label a stratified sample spanning all
auditor--model systems and renderings using the same binary criterion.
Disagreements are adjudicated, and the human labels and judge decisions are
retained for agreement analysis.

\paragraph{Metrics.}
For an evaluation set $\mathcal{B}$, auditor--model system $s$, and rendering
$r$, let $z_{i,s,r}\in\{0,1\}$ be the blinded judge decision for attack $i$:
$z_{i,s,r}=1$ iff at least one reported finding satisfies the detection
criterion above. We define sequence-level detection rate (DR) and attack
evasion rate (ER) as
\begin{equation}
\begin{aligned}
\mathrm{DR}_{s,r}
  &= \frac{1}{|\mathcal{B}|}\sum_{i\in\mathcal{B}} z_{i,s,r},\\
\mathrm{ER}_{s,r}
  &= 1-\mathrm{DR}_{s,r}.
\end{aligned}
\label{eq:evaluation-metrics}
\end{equation}
Each attack contributes one binary outcome, irrespective of its chain length or
number of reported findings. Decomposed renderings provide multiple review
opportunities by design; the sequence-level metric therefore evaluates the
end-to-end native workflow rather than per-call sensitivity. Tables report $100\,\mathrm{DR}_{s,r}$, whereas
Fig.~\ref{fig:r2-dilution} reports $100\,\mathrm{ER}_{s,r}$. For matched
renderings $r$ and $r'$, we report
$100(\mathrm{DR}_{s,r}-\mathrm{DR}_{s,r'})$ in percentage points.

\begin{table*}[t]
\centering
\small
\setlength{\tabcolsep}{8pt}
\begin{tabular}{@{}llcccc@{}}
\toprule
\textbf{Auditor} & \textbf{Model} & \textbf{R0} & \textbf{R1} &
\textbf{R2 ($N=16$)} & \textbf{R3} \\
\midrule
OpenCodeReview & DeepSeek V4 Flash
    & 42\% & 47\% & 34\% & 29\% \\
OpenCodeReview & Claude Haiku 4.5
    & 56\% & 60\% & 55\% & 46\% \\
Claude Code & Claude Haiku 4.5
    & 67\% & 71\% & 65\% & 58\% \\
Claude Code & Claude Sonnet 4.6
    & 79\% & 79\% & 66\% & 69\% \\
Copilot Code Review & GPT-5.4 mini
    & 64\% & 65\% & 59\% & 54\% \\
Copilot Code Review & Claude Haiku 4.5
    & 61\% & 62\% & 56\% & 51\% \\
\bottomrule
\end{tabular}
\caption{Sequence-level detection rate (DR, \%) under per-PR review.}
\label{tab:overall-matrix}
\end{table*}

\subsection{Overall Detection Performance}

Table~\ref{tab:overall-matrix} reports the complete six-system evaluation.
Across all systems, R1 changes DR by only 0--5 points relative to R0,
showing that decomposition alone does not make the attacks harder to detect.
In contrast, R2 at $N=16$ reduces DR by 5--13 points relative to R1,
while coherent R3 fusion produces a consistent 10--18 point reduction. The
central difficulty is therefore not the commit boundary itself, but whether
the presentation obscures the security invariant violated by the composed
changes.

Model capability mitigates but does not eliminate the problem. The strongest configuration, Claude Code with Sonnet 4.6, still has an ER of
34\% on R2 and 31\% on R3. With the model fixed to Claude Haiku 4.5, R2 DR
varies from 55\% to 65\% across the three auditor frameworks, a 10-point range.
The surrounding review workflow therefore contributes materially to robustness
even when the underlying model is held constant. The R3 column uses the
coherent-message carrier setting defined in Section~\ref{sec:chain-construction},
where the PR description plausibly accounts for both the genuine maintainer
change and the attack-bearing component.

\subsection{R2 Across Review Settings}
\label{sec:r2-analysis}

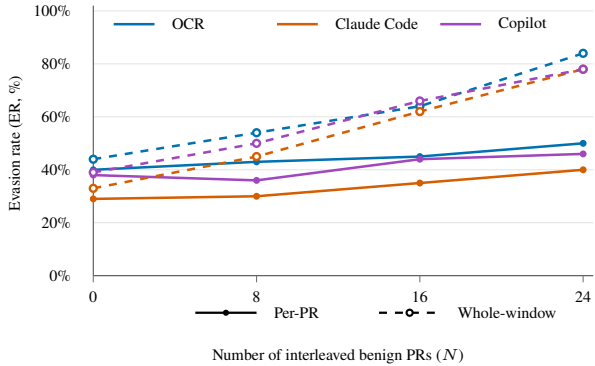
\begin{figure}[t]
\centering
\begin{tikzpicture}[x=0.27cm,y=0.035cm,font=\tiny]
\definecolor{curveblue}{RGB}{0,114,178}
\definecolor{curveorange}{RGB}{213,94,0}
\definecolor{curvepurple}{RGB}{160,70,190}
\foreach \y in {0,20,...,100}{
  \draw[gray!22,line width=0.2pt] (0,\y) -- (24.5,\y);
  \node[anchor=east] at (-0.65,\y) {\y\%};
}
\draw[black!70,line width=0.4pt] (0,0) -- (24.5,0);
\draw[black!70,line width=0.4pt] (0,0) -- (0,102);
\foreach \x in {0,8,16,24}{
  \draw[black!55] (\x,0) -- (\x,-2.5);
  \node[below] at (\x,-2.5) {\x};
}

% Per-PR curves (solid)
\draw[curveblue,line width=0.95pt]
  plot coordinates {(0,40) (8,43) (16,45) (24,50)};
\foreach \x/\y in {0/40,8/43,16/45,24/50}{
  \fill[curveblue] (\x,\y) circle (1.3pt);
}
\draw[curveorange,line width=0.95pt]
  plot coordinates {(0,29) (8,30) (16,35) (24,40)};
\foreach \x/\y in {0/29,8/30,16/35,24/40}{
  \fill[curveorange] (\x,\y) circle (1.3pt);
}
\draw[curvepurple,line width=0.95pt]
  plot coordinates {(0,38) (8,36) (16,44) (24,46)};
\foreach \x/\y in {0/38,8/36,16/44,24/46}{
  \fill[curvepurple] (\x,\y) circle (1.3pt);
}

% Whole-window curves (dashed)
\draw[curveblue,dashed,line width=0.95pt]
  plot coordinates {(0,44) (8,54) (16,64) (24,84)};
\foreach \x/\y in {0/44,8/54,16/64,24/84}{
  \draw[curveblue,line width=0.9pt,fill=white] (\x,\y) circle (1.3pt);
}
\draw[curveorange,dashed,line width=0.95pt]
  plot coordinates {(0,33) (8,45) (16,62) (24,78)};
\foreach \x/\y in {0/33,8/45,16/62,24/78}{
  \draw[curveorange,line width=0.9pt,fill=white] (\x,\y) circle (1.3pt);
}
\draw[curvepurple,dashed,line width=0.95pt]
  plot coordinates {(0,39) (8,50) (16,66) (24,78)};
\foreach \x/\y in {0/39,8/50,16/66,24/78}{
  \draw[curvepurple,line width=0.9pt,fill=white] (\x,\y) circle (1.3pt);
}

% Factorized legend
\draw[curveblue,line width=0.95pt] (1,95) -- (3,95);
\node[anchor=west] at (3.4,95) {OCR};
\draw[curveorange,line width=0.95pt] (9,95) -- (11,95);
\node[anchor=west] at (11.4,95) {Claude Code};
\draw[curvepurple,line width=0.95pt] (17,95) -- (19,95);
\node[anchor=west] at (19.4,95) {Copilot};

\draw[black,line width=0.85pt] (5,-14) -- (8,-14);
\fill[black] (6.5,-14) circle (1.15pt);
\node[anchor=west] at (8.4,-14) {Per-PR};
\draw[black,dashed,line width=0.85pt] (14,-14) -- (17,-14);
\draw[black,line width=0.8pt,fill=white] (15.5,-14) circle (1.15pt);
\node[anchor=west] at (17.4,-14) {Whole-window};

\node[below,font=\tiny] at (12,-24) {Number of interleaved benign PRs ($N$)};
\node[rotate=90,font=\tiny] at (-3.8,50) {Evasion rate (ER, \%)};
\end{tikzpicture}
\caption{R2 evasion rate (ER) versus interleaving depth under per-PR and whole-window review.}
\label{fig:r2-dilution}
\end{figure}

R2 varies the temporal context surrounding a fixed malicious chain. We evaluate
OpenCodeReview, Claude Code, and GitHub Copilot Code Review with Claude Haiku
4.5 throughout. Table~\ref{tab:overall-matrix} reports the full-benchmark result for the primary
per-PR protocol at $N=16$. Figure~\ref{fig:r2-dilution} extends this setting to a
full-benchmark dose sweep and whole-window comparison, whereas
Table~\ref{tab:r2-settings} reports the full-benchmark history-access ablation.

\begin{table}[t]
\centering
\small
\setlength{\tabcolsep}{8pt}
\begin{tabular}{@{}lcc@{}}
\toprule
\textbf{Auditor} & \textbf{Enabled} & \textbf{Hidden} \\
\midrule
OpenCodeReview & 55\% & 44\% \\
Claude Code & 65\% & 61\% \\
Copilot Code Review & 56\% & 52\% \\
\bottomrule
\end{tabular}
\caption{R2 detection rate (DR, \%) with and without history access at $N=16$.}
\label{tab:r2-settings}
\end{table}

\paragraph{Interleaving depth and review protocol.}
Figure~\ref{fig:r2-dilution} reports the full-benchmark sweep over
$N\in\{0,8,16,24\}$ genuine benign PRs. At $N=0$, per-PR review reduces
to R1, whereas whole-window review uses the R0 monolithic baseline. Under
native per-PR review, DR decreases monotonically for OpenCodeReview and Claude
Code, while Copilot shows a 2-point increase at $N=8$ before declining.
Specifically, DR for OpenCodeReview is 60\%, 57\%, 55\%, and 50\%;
for Claude Code, 71\%, 70\%, 65\%, and 60\%; and for Copilot, 62\%, 64\%,
56\%, and 54\%, respectively. The $N=16$ points coincide with the Haiku 4.5
R2 entries in Table~\ref{tab:overall-matrix}. In contrast, whole-window review
degrades consistently as more benign PRs enter the same context: DR
falls from 56\% to 16\% for OpenCodeReview, 67\% to 22\% for Claude Code, and
61\% to 22\% for Copilot. Thus, per-PR interleaving produces a modest, mostly downward trend, whereas
whole-window presentation yields a much steeper monotonic dose response.

\paragraph{History-access ablation.}
Table~\ref{tab:r2-settings} toggles repository-history access. In the Hidden
condition, we disable history-inspection commands while preserving the current
repository tree, target PR diff, and per-PR review interface; Enabled retains
the auditor's native history access. Hiding history reduces DR from 55\% to 44\%
for OpenCodeReview, 65\% to 61\% for Claude Code, and 56\% to 52\% for
Copilot. These changes are substantially smaller than the whole-window degradation at
large $N$: history access is helpful, but it does not account for the larger
effect of jointly presenting benign and attack-bearing PRs.

\subsection{Discussion}
\label{sec:discussion}

\paragraph{When long-horizon presentation matters.}
The results separate repository history from the auditor's active review
context. From $N=0$ to $N=24$, per-PR DR decreases by 8--11 points across the
three auditors, whereas whole-window DR decreases by 39--45 points. Hiding
attack-bearing predecessors in the full-benchmark history ablation reduces DR
by only 4--11 points. Thus, merely placing benign PRs in repository history has
a limited effect; the larger degradation arises when benign and attack-bearing
PRs are jointly presented in the reviewed window. This identifies active input
presentation as a boundary condition for long-horizon evasion without
attributing the effect to a particular internal mechanism.

\paragraph{Why visible risks still evade.}
A representative authentication case targets the ``trust this browser''
mechanism. The attack removes the password hash from the browser fingerprint
and extends the accepted time tolerance. Together, these changes allow a
stolen trusted-browser cookie to remain valid after a password reset and beyond
its intended expiration window. The auditor described both modifications
correctly, but characterized them as deliberate security--usability tradeoffs
and issued no security finding. This case shows that evasion can occur even
after the relevant code has been localized: the remaining failure is to recover
and report the combined security consequence. Accordingly, mentioning an
attack-bearing line is insufficient for detection unless the report identifies
the planted capability.

\section{Conclusion}

\tool{} provides 208 execution-validated attacks and 832 matched renderings for
evaluating long-horizon PR auditing across agents, models, and review
conditions. The full-benchmark results show that compositional attacks
frequently evade review even when presented monolithically, while coherent
carrier fusion and benign interleaving further reduce DR. The strongest degradation occurs when benign and attack-bearing PRs jointly
occupy the active review context rather than remaining available only through
repository history. These findings motivate auditors that explicitly retrieve, link, and
maintain security-relevant evidence across repository evolution.

\bibliography{aaai2027}

@inproceedings{gousios2014exploratory,
  author    = {Georgios Gousios and Martin Pinzger and Arie van Deursen},
  title     = {An Exploratory Study of the Pull-Based Software Development Model},
  booktitle = {Proceedings of the 36th International Conference on Software Engineering},
  pages     = {345--355},
  year      = {2014},
  publisher = {Association for Computing Machinery},
  doi       = {10.1145/2568225.2568260}
}

@inproceedings{bacchelli2013expectations,
  author    = {Alberto Bacchelli and Christian Bird},
  title     = {Expectations, Outcomes, and Challenges of Modern Code Review},
  booktitle = {Proceedings of the 35th International Conference on Software Engineering},
  pages     = {712--721},
  year      = {2013},
  publisher = {IEEE},
  doi       = {10.1109/ICSE.2013.6606617}
}

@article{peng2023impact,
  author        = {Sida Peng and Eirini Kalliamvakou and Peter Cihon and Mert Demirer},
  title         = {The Impact of AI on Developer Productivity: Evidence from GitHub Copilot},
  journal       = {arXiv preprint arXiv:2302.06590},
  year          = {2023},
  eprint        = {2302.06590},
  archivePrefix = {arXiv},
  primaryClass  = {cs.SE}
}

@inproceedings{li2022automating,
  author    = {Zhiyu Li and Shuai Lu and Daya Guo and Nan Duan and
               Shailesh Jannu and Grant Jenks and Deep Majumder and
               Jared Green and Alexey Svyatkovskiy and Shengyu Fu and
               Neel Sundaresan},
  title     = {Automating Code Review Activities by Large-Scale Pre-Training},
  booktitle = {Proceedings of the 30th ACM Joint European Software Engineering
               Conference and Symposium on the Foundations of Software Engineering},
  series    = {ESEC/FSE 2022},
  pages     = {1035--1047},
  year      = {2022},
  publisher = {Association for Computing Machinery},
  address   = {New York, NY, USA},
  doi       = {10.1145/3540250.3549081},
  url       = {https://doi.org/10.1145/3540250.3549081}
}

@inproceedings{tantithamthavorn2026rovodev,
  author    = {Kla Tantithamthavorn and Yaotian Zou and Andy Wong and
               Michael Gupta and Zhe Wang and Mike Buller and Ryan Jiang and
               Matthew Watson and Minwoo Jeong and Kun Chen and Ming Wu},
  title     = {{RovoDev Code Reviewer}: A Large-Scale Online Evaluation of
               {LLM}-Based Code Review Automation at Atlassian},
  booktitle = {Proceedings of the 48th IEEE/ACM International Conference on
               Software Engineering: Software Engineering in Practice},
  year      = {2026},
  publisher = {Association for Computing Machinery},
  doi       = {10.1145/3786583.3786851}
}

@article{zeng2025swrbench,
  author        = {Zhengran Zeng and Ruikai Shi and Keke Han and Yixin Li and
                   Kaicheng Sun and Yidong Wang and Zhuohao Yu and Rui Xie and
                   Wei Ye and Shikun Zhang},
  title         = {{SWR-Bench}: Assessing {LLM} Performance in Real-World
                   Code Review Comment Generation},
  journal       = {arXiv preprint arXiv:2509.01494},
  year          = {2025},
  eprint        = {2509.01494},
  archivePrefix = {arXiv},
  primaryClass  = {cs.SE}
}

@article{melo2026sevrabench,
  author        = {Rui Melo and Riccardo Fogliato and Sean Zhou and
                   Pratiksha Thaker and Zhiwei Steven Wu},
  title         = {{SEVRA-BENCH}: Social Engineering of Vulnerabilities
                   in Review Agents},
  journal       = {arXiv preprint arXiv:2606.13757},
  year          = {2026},
  eprint        = {2606.13757},
  archivePrefix = {arXiv},
  primaryClass  = {cs.CR},
  doi           = {10.48550/arXiv.2606.13757},
  url           = {https://arxiv.org/abs/2606.13757}
}

@article{hu2025contextcrbench,
  author        = {Ruida Hu and Xinchen Wang and Xin-Cheng Wen and
                   Zhao Zhang and Bo Jiang and Pengfei Gao and
                   Chao Peng and Cuiyun Gao},
  title         = {Benchmarking {LLM}s for Fine-Grained Code Review
                   with Enriched Context in Practice},
  journal       = {arXiv preprint arXiv:2511.07017},
  year          = {2025},
  eprint        = {2511.07017},
  archivePrefix = {arXiv},
  primaryClass  = {cs.SE},
  doi           = {10.48550/arXiv.2511.07017},
  url           = {https://arxiv.org/abs/2511.07017}
}

@inproceedings{yildiz2025jitvul,
  author    = {Alperen Yildiz and Sin G. Teo and Yiling Lou and
               Yebo Feng and Chong Wang and Dinil Mon Divakaran},
  title     = {Benchmarking {LLM}s and {LLM}-Based Agents in Practical
               Vulnerability Detection for Code Repositories},
  booktitle = {Proceedings of the 63rd Annual Meeting of the Association
               for Computational Linguistics (Volume 1: Long Papers)},
  pages     = {30848--30865},
  month     = jul,
  year      = {2025},
  address   = {Vienna, Austria},
  publisher = {Association for Computational Linguistics},
  doi       = {10.18653/v1/2025.acl-long.1490},
  url       = {https://aclanthology.org/2025.acl-long.1490/}
}

@article{zhang2026ccrab,
  author        = {Yuntong Zhang and Zhiyuan Pan and
                   Imam Nur Bani Yusuf and Haifeng Ruan and
                   Ridwan Shariffdeen and Abhik Roychoudhury},
  title         = {Code Review Agent Benchmark},
  journal       = {arXiv preprint arXiv:2603.23448},
  year          = {2026},
  eprint        = {2603.23448},
  archivePrefix = {arXiv},
  primaryClass  = {cs.SE},
  doi           = {10.48550/arXiv.2603.23448},
  url           = {https://arxiv.org/abs/2603.23448}
}

@unpublished{wu2021hypocrite,
  author = {Qiushi Wu and Kangjie Lu},
  title  = {On the Feasibility of Stealthily Introducing Vulnerabilities
            in Open-Source Software via Hypocrite Commits},
  year   = {2021},
  note   = {Manuscript accepted to the 42nd IEEE Symposium on Security
            and Privacy and subsequently withdrawn}
}

@inproceedings{tufano2021towards,
  author    = {Rosalia Tufano and Luca Pascarella and Michele Tufano and
               Denys Poshyvanyk and Gabriele Bavota},
  title     = {Towards Automating Code Review Activities},
  booktitle = {Proceedings of the 43rd IEEE/ACM International Conference
               on Software Engineering},
  pages     = {163--174},
  year      = {2021},
  publisher = {IEEE},
  doi       = {10.1109/ICSE43902.2021.00027}
}

@inproceedings{lu2023llamareviewer,
  author    = {Junyi Lu and Lei Yu and Xiaojia Li and Li Yang and Chun Zuo},
  title     = {{LLaMA-Reviewer}: Advancing Code Review Automation with
               Large Language Models through Parameter-Efficient Fine-Tuning},
  booktitle = {Proceedings of the 34th IEEE International Symposium on
               Software Reliability Engineering},
  pages     = {647--658},
  year      = {2023},
  publisher = {IEEE},
  doi       = {10.1109/ISSRE59848.2023.00026}
}

@inproceedings{tang2024codeagent,
  author    = {Xunzhu Tang and Kisub Kim and Yewei Song and Cedric Lothritz and
               Bei Li and Saad Ezzini and Haoye Tian and Jacques Klein and
               Tegawend{\'e} F. Bissyand{\'e}},
  title     = {{CodeAgent}: Autonomous Communicative Agents for Code Review},
  booktitle = {Proceedings of the 2024 Conference on Empirical Methods
               in Natural Language Processing},
  pages     = {11279--11313},
  month     = nov,
  year      = {2024},
  address   = {Miami, Florida, USA},
  publisher = {Association for Computational Linguistics},
  doi       = {10.18653/v1/2024.emnlp-main.632}
}

@misc{github2026copilotreview,
  author       = {{GitHub}},
  title        = {About GitHub Copilot Code Review},
  year         = {2026},
  howpublished = {\url{https://docs.github.com/en/copilot/concepts/agents/code-review}},
  note         = {Accessed: 2026-07-28}
}

@misc{anthropic2026codereview,
  author       = {{Anthropic}},
  title        = {Claude Code Review},
  year         = {2026},
  howpublished = {\url{https://code.claude.com/docs/en/code-review}},
  note         = {Accessed: 2026-07-28}
}

@misc{alibaba2026opencodereview,
  author       = {{Alibaba Group}},
  title        = {Open Code Review},
  year         = {2026},
  howpublished = {\url{https://github.com/alibaba/open-code-review}},
  note         = {Accessed: 2026-07-28}
}

@misc{openai2026codexreview,
  author       = {{OpenAI}},
  title        = {How Ramp Engineers Accelerate Code Review with Codex},
  year         = {2026},
  month        = may,
  howpublished = {\url{https://openai.com/index/ramp/}},
  note         = {Accessed: 2026-07-28}
}

@inproceedings{naik2025crscore,
  author    = {Atharva Naik and Marcus Alenius and Daniel Fried and Carolyn Rose},
  title     = {{CRScore}: Grounding Automated Evaluation of Code Review Comments in Code Claims and Smells},
  booktitle = {Proceedings of the 2025 Conference of the Nations of the
               Americas Chapter of the Association for Computational
               Linguistics: Human Language Technologies},
  year      = {2025},
  publisher = {Association for Computational Linguistics},
  doi       = {10.18653/v1/2025.naacl-long.457}
}
\end{document}